# Evaluation of ChatGPT-Generated Medical Responses: A Systematic Review and Meta-Analysis

Qiuhong Wei,[*a,f,g] Zhengxiong Yao,[*b] Ying Cui,[*c] Bo Wei,[d] Zhezhen Jin,[#e] and Ximing Xu[#a]

a. Big Data Center for Children's Medical Care, Children's Hospital of Chongqing Medical University, Chongqing, China;

b. Department of Neurology, Children's Hospital of Chongqing Medical University, Chongqing, China;

c. Department of Biomedical Data Science, Stanford University School of Medicine, Stanford, CA, USA;

d. Department of Global Statistics and Data Science, BeiGene USA Inc., San Mateo, CA, USA;

e. Department of Biostatistics, Mailman School of Public Health, Columbia University, New York, NY, USA;

f. Children Nutrition Research Center, Children's Hospital of Chongqing Medical University, Chongqing, China;

g. National Clinical Research Center for Child Health and Disorders, Ministry of Education Key Laboratory of Child Development and Disorders, China International Science and Technology Cooperation Base of Child Development and Critical Disorders, Chongqing Key Laboratory of Childhood Nutrition and Health, Chongqing, China.

* The authors made an equal contribution

[#] Corresponding author: Zhezhen Jin and Ximing Xu share the corresponding authorship

Zhezhen Jin, Prof, Ph.D., Department of Biostatistics, Mailman School of Public Health, Columbia University, 722 West 168th Street New York, NY, 10032, USA, zj7@cumc.columbia.edu.

Ximing Xu, Prof, Ph.D., Big Data Center for Children's Medical Care, Children's Hospital of Chongqing Medical University, Chongqing, China, No 136. Zhongshan 2nd Rd, Yuzhong District, Chongqing, 400014, China, 86-18526547007, ximing@hospital.cqmu.edu.cn.

**Abstract**

Large language models such as ChatGPT are increasingly explored in medical domains. However, the absence of standard guidelines for performance evaluation has led to methodological inconsistencies. This study aims to summarize the available evidence on evaluating ChatGPT's performance in medicine and provide direction for future research. We searched ten medical literature databases on June 15, 2023, using the keyword "ChatGPT". A total of 3520 articles were identified, of which 60 were reviewed and summarized in this paper and 17 were included in the meta-analysis. The analysis showed that ChatGPT displayed an overall integrated accuracy of 56% (95% CI: 51%-60%, $I^2 = 87\%$) in addressing medical queries. However, the studies varied in question resource, question-asking process, and evaluation metrics. Moreover, many studies failed to report methodological details, including the version of ChatGPT and whether each question was used independently or repeatedly. Our findings revealed that although ChatGPT demonstrated considerable potential for application in healthcare, the heterogeneity of the studies and insufficient reporting may affect the reliability of these results. Further well-designed studies with comprehensive and transparent reporting are needed to evaluate ChatGPT's performance in medicine.

**Introduction**

In recent years, large language models (LLMs) have attracted attention for their potential to improve traditional approaches in diverse domains[1]. Within healthcare, ChatGPT is a notable example that exhibits promising characteristics in generating text that resembles human-like communication[2]. These features have led to exploratory uses of ChatGPT in tasks such as responding to medical inquiries and crafting detailed medical content. While there is growing interest in ChatGPT's ability to contribute to medical education, enhance scientific writing, and possibly support clinical decision-making, it also raises various concerns that warrant careful evaluation by practitioners and researchers [3–6].

Since the launch of ChatGPT, a generative artificial intelligence, numerous peer-reviewed scientific papers have been published on its medical applications. ChatGPT's performance has been evaluated across a wide range of medical fields, including basic disciplines like biochemistry[7], microbiology[8], and genetics[9], as well as specialized areas such as internal medicine (e.g., gastroenterology[10], cardiology[11], infectious diseases[12], and endocrinology[13]), surgical medicine (e.g., hepatobiliary surgery[14], orthopedics[15], urology[16], and neurosurgery[17]), obstetrics and gynecology[18], and radiology and laboratory medicine[19–20]. ChatGPT has generated considerable attention by achieving accuracies ranging from 36% to 90%. However, it is crucial to use these findings with caution. The field is still developing and lacks standardized guidelines for the evaluation of LLMs' performances, which leads to inconsistencies across studies. Such discrepancies raise concerns about the sources of the questions/queries, the question-asking process, and the evaluation metrics. Given the growing implementation of LLMs in medicine, a systematic review of the approaches used for the evaluation of the performance of LLMs becomes essential.

Recently, several studies reviewed the advancement in LLMs and evaluated LLM performance in general [21–22]. In the specific area of healthcare, there exist several overviews on the application of LLM in education, research, practice, scientific writing, and ethical considerations[22–26], but there is a noticeable gap in the literature regarding evaluations of LLMs' performance in medical questions/queries designed by researchers, particularly in the use of

ChatGPT. It is necessary to evaluate and examine the variations in research design, question/query formulation, and evaluation metrics across different studies. In this study, we conduct a systematic review of the existing literature evaluating LLMs in the use of ChatGPT. Our objectives are to: (a) review the available evidence on ChatGPT's performance in answering medical questions, (b) synthesize these findings through meta-analysis, and (c) examine the methodologies used in the literature. Our study provides insights into the current state of research and useful information for future studies.

## Methods

This review and meta-analysis were conducted following the guidelines of the Preferred Reporting Items for Systematic Reviews and Meta-Analyses (PRISMA)[27]. The study is registered with PROSPERO (CRD42023456327).

Eligibility criteria

To ensure consistency across the studies included in this analysis, the following selection criteria were applied:

(a) Inclusion criteria:

i. Evaluation of ChatGPT's performance in medical domains;

ii. ChatGPT responded to medical questions.

(b) Exclusion criteria:

i. Review articles, comments, and patents;

ii. Evaluation of ChatGPT's writing proficiency or other non-medicine-related performances;

iii. Studies in preprint format.

Data Source and Search Strategy

An exhaustive literature search was carried out on June 15, 2023, across the following databases: Medline, Embase, Web of Science, Scopus, Cochrane, IEEE Xplore, Wan Fang Database, VIP Database for Chinese Technical Periodicals (VIP), Chinese BioMedical Literature Database (CBM), and China National Knowledge Infrastructure (CNKI). To retrieve a broad spectrum of relevant literature, we used the keyword "ChatGPT" as a search term, without restrictions on publication type, language, or publication date. Two authors (WQH and YZX) independently performed the database search and imported all records into the NoteExpress software.

Study Selection

Duplicate records were first removed based on title, author, and publication date. Titles, abstracts, and keywords were examined for relevance. Full texts of identified studies were screened independently for inclusion. Any disagreements were addressed through discussion between WQH and YZX.

Data Extraction

A data extraction form was collaboratively designed and utilized by two authors (WQH and YZX). The extraction of data from full texts was conducted independently by the two authors. A third reviewer (XXM) was consulted to resolve conflicts. Studies meeting the criteria were included in the systematic review and meta-analysis. General characteristics data included author, publication year, region, publication type, medical disciplines, and specific features concerning the design of medical knowledge assessment for ChatGPT.

Specifically, the source of questions posed to ChatGPT was extracted, including various examination question banks, author-designed questions (typically inspired by clinical vignettes or physicians' clinical experience), and open websites like Facebook, medical knowledge forums, and patient-physician Q&A forums. Additional extracted data included the question source's publication date, the language of the question, the number of questions, and the question type (such as multiple-choice and open-ended questions). In addition, the characters related to conversation processes were considered: ChatGPT version, mode of inquiry (via web or Application Programming Interface [API]), whether questions were posed independently or repeatedly, and whether designed prompts like "Please act as a medical student taking the medical exam" were used. Assessment-related information was also extracted, including the rater, assessment metrics, and performance of ChatGPT.

Quality assessment

The methodological quality (risk of bias) of each included study was independently evaluated by WQH and YZX according to Quality Assessment of Diagnostic Accuracy Studies-2 (QUADAS-2)[28]. This instrument assessed various domains including the selection of questions posed to LLMs, the index test, the reference standard, and flow and timing.

Evidence Synthesis and Statistic Analysis

Studies that evaluated the accuracy of ChatGPT by using objective questions within the medical knowledge domain were integrated into a meta-analysis to determine its overall performance. To assess the potential for publication bias, we applied both visual and statistical methods: funnel plots and Egger's regression test. The I2 statistic was employed to assess the effect of heterogeneity on the pooled results. Random-effect models were used when there was significant heterogeneity (I2 > 50%). Otherwise, fixed-effect models were used. Subgroup analysis was carried out to investigate the potential sources of heterogeneity and compare performances across different subgroups. A sensitivity analysis was carried out to assess the robustness of the results. Accuracy was reported with 95% confidence intervals (CIs). The p-value 0.05 was used as the significance level. The "meta" package in R 4.1.0 was utilized for the meta-analysis and subgroup analysis, sensitivity analysis, and publication bias.

## Results

Literature Screening and Selection

A total of 3520 records were identified through searches in various databases. After removing duplicates, 771 studies were retrieved from ten different databases. A thorough review was conducted based on predefined inclusion and exclusion criteria, which narrowed down to 60 studies for systematic review [5, 7–8, 11, 14–15, 29–82] (Supplementary Table 1). Of these, 17 were further examined with a formal meta-analysis[15, 67–82]. No publication bias was detected among the included studies, as indicated by Egger's test, which yielded a t-value of 0.02 with a p-value of 0.984 (Supplementary Figure 1 for funnel plots). The inter-rater agreement for study inclusion and data extraction between WQH and YZX was 58/60 (97%), with the involvement of a third reviewer for 2/60 (3%) studies. The detailed process of literature screening and selection is depicted in Figure 1.

Characteristics of Included Studies

The 60 studies were published between January and May 2023 and encompassed various publication types and medical disciplines. Specifically, the publication consisted of 37 articles, 16 letters, and seven short reports including correspondences, brief communications, and brief reports. In terms of the medical disciplines, the studies had 16 in internal medicine, 15 in surgery, two in obstetrics and gynecology, 7 in specialty departments (such as otolaryngology and ophthalmology), 11 in clinical auxiliary departments, and 9 in other areas, such as medical license examinations and basic medical disciplines. Additional details of these characteristics can be found in Figure 2.

Design of Medical Examinations and Assessments

*Question/Prompt Sources*

The sources of questions posed to ChatGPT varied. Twenty-two studies used author-designed questions, 21 used questions from examination banks, 13 used questions from open websites, 2 studies used existing questionnaires, one study used published literature, and one study used questions from various sources including textbooks, examination question banks, and clinical cases. Among the 38 studies that used publicly available sources

(i.e., those that used non-author-designed questions), 13 studies selected questions developed after September 2021, 7 selected questions before September 2021, 7 selected questions both before and after September 2021, and 11 did not report the timing. Regarding language, 37 studies used questions in English, one in both English and Chinese, one in Dutch, and one in Korean; 20 studies did not report the language used. Regarding the number of questions, 57 studies reported counts of questions, with a median and interquartile range of 66 (25, 195), 2 studies reported the use of seven and eight sets of questions without specifying the exact number, and one did not report the number of questions used. In terms of the type of questions, 19 studies used multiple-choice questions, 36 studies used open-ended questions, and 5 studies used both types.

*Conversation Process*

In the conversation process with ChatGPT, different studies employed various versions and modes of inquiry. Among the included studies, 13 utilized GPT-3.5, 3 opted for GPT-3, 4 leveraged GPT-4, while some combined different versions (3 studies used both GPT-4 and GPT-3.5, and 1 study used both GPT-3 and GPT-4). Interestingly, 16 studies mentioned the date of usage without specifying the version, and 20 did not provide information on either aspect. As for the mode of inquiry, 43 studies interacted with ChatGPT via a web-based interface, none reported using an API, and 17 did not report the mode of accessing GPT. There was also variation in the question-posing process: 18 studies asked questions independently, with a new chat window for each, and only one study entered all questions in a single session; the remaining 41 did not disclose any related information. Repetition of inquiry was another aspect: 14 studies asked each question only once, others repeated questions multiple times (6 studies repeated twice, 7 repeated three times, 3 repeated five times, and 1 repeated six times), and 29 studies did not report on repetition. The studies also differed in their approach on prompting. One study provided a specifically designed prompt that guided ChatGPT to act as a laboratory medicine expert ("Act as a personal assistant who is a laboratory medicine expert and can interpret lab exam results and help patients understand them."), 40 studies simply prompted ChatGPT to answer the question, and 19 studies did not report this aspect of the interaction.

*Evaluation of ChatGPT's Performance*

The evaluation of ChatGPT's performance across the included studies varied in both the raters and the metrics used. Among the studies, 39 engaged medical professionals to evaluate ChatGPT's responses, most often involving two or three raters (26 out of 39 studies). One study fed ChatGPT's answers directly into an automated scoring system within the examination platform. The remaining 20 studies did not report information about the evaluators. In terms of assessment metrics, the most commonly used was accuracy, with a small number of studies also considering completeness, safety, appropriateness, readability, humanistic care, word count, and inter-rater consistency (for studies with multiple raters). The majority of the studies (57 out of 60) quantified ChatGPT's performance, employing a range of quantitative metrics such as accuracy rate, Likert scale, and various scoring systems.

Study Quality Evaluation

The methodological quality (risk of bias) of each included study was independently evaluated using Quality Assessment of Diagnostic Accuracy Studies-2 (QUADAS-2). In the quality evaluation, more than half of the studies had low risk in several domains. Specifically, 37 out of 60 studies showed a low risk in the domain of prompt selection, 57 out of 60 were in the standard reference domain, and all 60 studies had a low risk in the flow and timing domains. On the other hand, 30 studies were identified with an unclear risk of bias in the index test domain (Supplementary Figure 2a). Of the 17 studies included in the meta-analysis, 13 had low risk across all domains, the remaining four studies demonstrated low risk in the selection of prompt, standard reference, and flow and timing, but presented a risk of bias in the index test (Supplementary Figure 2b).

Meta-analysis

ChatGPT, as a product of general-purpose artificial intelligence with extensive training in human knowledge, has great potential to be used in medical fields for various disciplines, populations, and languages. The diversity of questions and evaluation methods used in existing studies underscores the complexity of assessing ChatGPT's performance. A meta-analysis offers a way of synthesizing these diverse findings, overcoming the limitations of individual studies, and providing a more comprehensive understanding of ChatGPT's ability to answer medical-related questions.

To ensure the comparability of the studies included in the meta-analysis, we carefully searched the literature: among the 60 studies included in the systematic review, 41 posed open-ended questions to ChatGPT. In the 41 studies, medical professionals evaluated ChatGPT's responses using various scales, such as Likert-5, Likert-7, scores of 0-2, and scores of 0-6, along with different performance metrics, including reliability, appropriateness, safety, and accuracy. The diversity of evaluation methods made it challenging to identify a standard quantitative metric to uniformly assess ChatGPT's performance across these studies. The remaining 19 studies utilized objective, multiple-choice questionnaires. Among the 19 studies, two did not report the specific number of correct responses. The remaining 17 studies reported both the total number of questions and the number of correct answers provided by ChatGPT. As a result, the 17 studies were included in a formal meta-analysis. As illustrated in Figure 3, ChatGPT achieved an integrated accuracy of 56% [95% CI: 51%-60%] using a random-effect model ($I^2 = 87\%$).

Subgroup Analysis and Meta-regression

We further examined the potential sources of heterogeneity with subgroup analysis and meta-regression. The subgroup analysis is based on the date of question origin (before or after the cutoff date for GPT's knowledge database), GPT version (GPT-3, GPT-3.5, or GPT-4), major department categories (internal medicine, surgery, and auxiliary departments), question repetition (repeat or not repeat), and study quality (no risk of bias or exiting risk of bias). The results of the subgroup analysis showed that ChatGPT achieved an accuracy rate of 63% [95%CI: 60%-65%] in internal medicine and 49% [95%CI: 44%- 54%] in surgery, with no significant differences observed in other categories. In the meta-regression analysis, the "Internal Medicine" category exhibited a significant effect on the accuracy of ChatGPT, as indicated by an estimated regression coefficient of 0.67 (P-value < 0.01). Other subgroup analyses did not show statistically significant effects on accuracy(all P-values > 0.05).

Sensitivity Analysis

To evaluate the robustness of our meta-analysis results, we conducted a sensitivity analysis using the leave-one-out method, where the meta-analysis was repeated by excluding one study at a time. The sensitivity analysis showed that no individual study had an excessive impact on the overall meta-analysis results (Supplementary Figure 3), which confirms the robustness of our results.

**Discussion**

This comprehensive systematic review and meta-analysis provided the first extensive assessment of the performance of ChatGPT in medical question-answering. After analyzing multiple medical studies across different disciplines, the result suggests that ChatGPT displays an overall integrated accuracy of 56% in addressing medical queries. However, the heterogeneity of the studies and insufficient reporting may affect the reliability of these results. Further well-designed studies with comprehensive and transparent reporting are needed to evaluate ChatGPT's performance in medicine.

In the subgroup analysis and meta-regression, ChatGPT achieved higher accuracy in internal medicine (63%) than in surgery (49%). This discrepancy might be rooted in the inherent differences between these two disciplines and how they align with ChatGPT's strengths and limitations. Internal medicine often involves the diagnosis of diseases based on a synthesis of symptoms, laboratory results, and existing medical literature [83]. This process requires extensive knowledge, logical reasoning, and the ability to correlate various pieces of information, areas where AI models like ChatGPT excel. With access to vast medical databases, ChatGPT can efficiently analyze and interpret complex medical data, thereby assisting the diagnostic process. Conversely, surgical medicine emphasizes manual skills, tactile sensations, and real-world experience[84]. While ChatGPT can access surgical guidelines, it lacks the ability to comprehend the nuances of surgical procedures and the tactile feedback necessary for surgical decision-making. The complexity of surgical scenarios may also require an understanding that goes beyond text-based AI models, leading to less accurate responses. This difference in performance between internal medicine and surgery highlights the need for careful consideration when employing AI models across medical specialties, recognizing that their effectiveness may vary.

Besides, although ChatGPT showed potential in responding to medical queries, it is important to interpret these results with caution. Many studies included in this review employed open-ended questions, evaluated by clinicians with different scales and metrics, possibly introducing a source of bias. Moreover, the lack of clarity on how ChatGPT processes prompt and generates responses makes it challenging to determine the reliability and repeatability of its answers. Our findings revealed that the reporting quality of the included studies showed notable variation, and many studies did not report important methodological details, such as the version of ChatGPT used, the repetition of the inquiry, and whether each question was posed independently. These omissions indicate a pressing need for standardized reporting guidelines for evaluating LLMs in medical domains.

This review utilized the QUADAS-2 to evaluate the methodological quality of the included studies, recognizing that there are currently no tools specifically tailored for assessing the quality of Large Language Model (LLM) research. Traditional instruments such as the PROBAST for prediction model studies[85], the Cochrane Risk of Bias for Randomized Controlled Trials[86], the Methodological Index for Non-Randomized Studies (MINORS)[87], and the Newcastle-Ottawa Scale (NOS)[88] fall short due to differences in study design, rendering the inappropriateness for this context. In contrast, QUADAS-2, grounded in evidence-based principles for the assessment of diagnostic accuracy studies, emerged as a more appropriate choice. The rationale for this selection lies in the parallels between evaluating ChatGPT's performance and diagnostic accuracy study designs, particularly in areas such as case selection and reference standard determination.

However, it's worth acknowledging that QUADAS-2 has its limitations in the context of LLM evaluation. Some items in QUADAS-2, such as the consideration of an appropriate interval between the test and the reference standard, may not be fully applicable. Additionally, the tool does not cover the components like the process of posing questions and the specific metrics of evaluation. Recognizing these gaps, we have proposed a comprehensive checklist to appraise the performance of large language models in medicine, supplementing the qualitative framework advanced by Howard et al. This checklist is organized into four sections encompassing 20 items: (a) task generation; (b) LLM version; (c) conversation structure; and (d) evaluation procedure (Table 1). By offering a structured and detailed approach, the checklist aims to facilitate a rigorous scientific assessment of large language models in the medical field. In doing so, it lays a foundation for the ongoing refinement and progression of these innovative tools, aligning with the evolving needs and standards in medical practice.

There are several limitations worth considering. First, although we extensively searched ten databases, there may still be relevant studies not included, potentially affecting the completeness of the review. Second, the rigorous selection criteria employed in this study led to a limited number of studies included in the meta-analysis, thereby constraining the generalizability of the findings. Third, the presence of high inter-study heterogeneity coupled with insufficient reporting in some studies has the potential to affect the strength of our conclusions. Fourth, a lack of representation in certain fields, including pediatrics, means that ChatGPT's capabilities in answering medical questions related to these specialties have not been fully explored. This gap not only limits our understanding of the AI's overall proficiency in the medical field but also highlights areas for future investigation. To enhance the reliability and validity of subsequent evaluations, the development of standardized guidelines for implementing and reporting assessments is highly recommended. Such guidelines would foster consistency in evaluation methodologies, allowing for more accurate and unbiased comparisons across studies, and thus contributing to a more comprehensive understanding of the potential applications of AI language models like ChatGPT in healthcare.

In conclusion, this study provides the first comprehensive assessment of ChatGPT's capabilities in answering medical questions. Despite significant heterogeneity among the included studies, ChatGPT demonstrates considerable potential for application in healthcare. Future research should focus on standardizing evaluation methods, investigating the factors that influence performance, and exploring strategies for effectively integrating AI language models into healthcare systems.

**Author Contributions**

Qiuhong Wei: Conceptualization, Methodology, Software, Validation, Formal analysis, Resources, Writing-Original Draft, and Writing-Review and Editing. Zhengxiong Yao: Methodology, Software, Validation, Formal Analysis, Writing-Original Draft, and Writing-Review and Editing. Ying Cui: Conceptualization, Methodology, Validation, Formal Analysis, Resources, and Writing-Review and Editing Investigation. Bo Wei: Methodology, Software, Validation, Writing-Review and Editing. Zhezhen Jin: Conceptualization, Methodology, Software, Validation, Formal Analysis, Resources, and Writing-Review and Editing. Ximing Xu: Conceptualization, Methodology, Software, Validation, Formal analysis, Resources, Writing-Review and Editing.

**Declaration of competing interest**

All authors declare no conflicts of interest

**Acknowledgments**

Not applicable

**Funding**

Not applicable

**Data Sharing**

The extracted data contributing to the meta-analysis are available in the Supplementary Material. Any additional data are available on request.

# References:

1. Bi K, Xie L, Zhang H, Chen X, Gu X, Tian Q. Accurate medium-range global weather forecasting with 3D neural networks. *Nature* 2023;619(7970):533-538.doi: 10.1038/s41586-023-06185-3.PubMed: 37407823
2. Ray PP. ChatGPT: A comprehensive review on background, applications, key challenges, bias, ethics, limitations and future scope. *Internet of Things and Cyber-Physical Systems* 2023;3:121-154.doi: 10.1016/j.iotcps.2023.04.003
3. Huang J, Tan M. The role of ChatGPT in scientific communication: writing better scientific review articles. *Am J Cancer Res* 2023;13(4):1148-1154.PubMed: 37168339
4. Lee H. The rise of ChatGPT: Exploring its potential in medical education. *Anat Sci Educ* 2023.doi: 10.1002/ase.2270
5. Liu SR, Wright AP, Patterson BL, et al. Using AI-generated suggestions from ChatGPT to optimize clinical decision support. *J Am Med Inform Assn* 2023.doi: 10.1093/jamia/ocad072
6. Singhal K, Azizi S, Tu T, et al. Large language models encode clinical knowledge. *Nature* 2023.doi: 10.1038/s41586-023-06291-2.PubMed: 37438534
7. Ghosh A, Bir A. Evaluating ChatGPT's Ability to Solve Higher-Order Questions on the Competency-Based Medical Education Curriculum in Medical Biochemistry. *CUREUS JOURNAL OF MEDICAL SCIENCE* 2023;15(4).doi: 10.7759/cureus.37023
8. Das D, Kumar N, Longjam LA, et al. Assessing the Capability of ChatGPT in Answering First- and Second- Order Knowledge Questions on Microbiology as per Competency- Based Medical Education Curriculum. *CUREUS JOURNAL OF MEDICAL SCIENCE* 2023;15(3).doi: 10.7759/cureus.36034
9. Duong D, Solomon BD. Analysis of large-language model versus human performance for genetics questions, 2023.
10. Yeo YH, Samaan JS, Ng WH, et al. Assessing the performance of ChatGPT in answering questions regarding cirrhosis and hepatocellular carcinoma. *Clinical and molecular hepatology* 2023.doi: 10.3350/cmh.2023.0089
11. Sarraju A, Bruemmer D, Van Iterson E, Cho L, Rodriguez F, Laffin L. Appropriateness of Cardiovascular Disease Prevention Recommendations Obtained From a Popular Online Chat-Based Artificial Intelligence Model. *JAMA* 2023;329(10):842-844.doi: 10.1001/jama.2023.1044.PubMed: 36735264
12. Howard A, Hope W, Gerada A. ChatGPT and antimicrobial advice: the end of the consulting infection doctor? *Lancet Infect Dis* 2023.doi: 10.1016/S1473-3099(23)00113-5.PubMed: 36822213
13. Nakhleh A, Spitzer S, Shehadeh N. ChatGPT's Response to the Diabetes Knowledge Questionnaire: Implications for Diabetes Education. *Diabetes Technol the* 2023.doi: 10.1089/dia.2023.0134
14. Cao JJ, Kwon DH, Ghaziani TT, et al. Accuracy of Information Provided by ChatGPT Regarding Liver Cancer Surveillance and Diagnosis. *AJR Am J Roentgenol* 2023.doi: 10.2214/AJR.23.29493.PubMed: 37222278
15. Lum ZC. Can Artificial Intelligence Pass the American Board of Orthopaedic Surgery Examination? Orthopaedic Residents Versus ChatGPT. *Clin Orthop Relat Res* 2023.doi: 10.1097/CORR.0000000000002704.PubMed: 37220190
16. Zhu L, Mou W, Chen R. Can the ChatGPT and other Large Language Models with internet-connected database solve the questions and concerns of patient with prostate cancer? 2023.
17. Hopkins BS, Nguyen VN, Dallas J, et al. ChatGPT versus the neurosurgical written boards: a comparative analysis of artificial intelligence/machine learning performance on neurosurgical board-style questions. *J Neurosurg* 2023:1-8.doi: 10.3171/2023.2.JNS23419

18. Li SW, Kemp MW, Logan SJS, et al. ChatGPT Outscored Human Candidates in a Virtual Objective Structured Clinical Examination (OSCE) in Obstetrics and Gynecology. *Am J Obstet Gynecol* 2023.doi: 10.1016/j.ajog.2023.04.020
19. Bhayana R, Krishna S, Bleakney RR. Performance of ChatGPT on a Radiology Board-style Examination: Insights into Current Strengths and Limitations. *Radiology* 2023:230582.doi: 10.1148/radiol.230582
20. Johnson D, Goodman R, Patrinely J, et al. Assessing the Accuracy and Reliability of AI-Generated Medical Responses: An Evaluation of the Chat-GPT Model. *Res Sq* 2023.doi: 10.21203/rs.3.rs-2566942/v1.PubMed: 36909565
21. Chang Y, Wang X, Wang J, et al. A survey on evaluation of large language models. *arXiv preprint arXiv:2307.03109* 2023
22. Thirunavukarasu AJ, Ting D, Elangovan K, Gutierrez L, Tan TF, Ting D. Large language models in medicine. *Nat Med* 2023.doi: 10.1038/s41591-023-02448-8.PubMed: 37460753
23. Sallam M. ChatGPT Utility in Healthcare Education, Research, and Practice: Systematic Review on the Promising Perspectives and Valid Concerns. *HEALTHCARE* 2023;11(6).doi: 10.3390/healthcare11060887
24. Vaishya R, Misra A, Vaish A. ChatGPT: Is this version good for healthcare and research? *Diabetes and Metabolic Syndrome: Clinical Research and Reviews* 2023;17(4).doi: 10.1016/j.dsx.2023.102744
25. Li J, Dada A, Kleesiek J, Egger J. ChatGPT in Healthcare: A Taxonomy and Systematic Review, 2023.
26. Harrer S. Attention is not all you need: the complicated case of ethically using large language models in healthcare and medicine. *Ebiomedicine* 2023;90:104512.doi: 10.1016/j.ebiom.2023.104512.PubMed: 36924620
27. McInnes M, Moher D, Thombs BD, et al. Preferred Reporting Items for a Systematic Review and Meta-analysis of Diagnostic Test Accuracy Studies: The PRISMA-DTA Statement. *JAMA* 2018;319(4):388-396.doi: 10.1001/jama.2017.19163.PubMed: 29362800
28. Whiting PF, Rutjes AW, Westwood ME, et al. QUADAS-2: a revised tool for the quality assessment of diagnostic accuracy studies. *Ann Intern Med* 2011;155(8):529-36.doi: 10.7326/0003-4819-155-8-201110180-00009.PubMed: 22007046
29. Sarink M, Bakker IL, Anas AA, Yusuf E. A Study on the Performance of ChatGPT in Infectious Diseases Clinical Consultation. *Clinical microbiology and infection : the official publication of the European Society of Clinical Microbiology and Infectious Diseases* 2023.doi: 10.1016/j.cmi.2023.05.017
30. Wagner MW, Ertl-Wagner BB. Accuracy of Information and References Using ChatGPT-3 for Retrieval of Clinical Radiological Information. *Canadian Association of Radiologists Journal* 2023.doi: 10.1177/08465371231171125
31. Xie Y, Seth I, Hunter-Smith DJ, Rozen WM, Ross R, Lee MT. Aesthetic Surgery Advice and Counseling from Artificial Intelligence: A Rhinoplasty Consultation with ChatGPT. *Aesthet Plast Surg* 2023.doi: 10.1007/s00266-023-03338-7
32. Sivasubramanian J, Shaik Hussain SM, Virudhunagar Muthuprakash S, Periadurai ND, Mohanram K, Surapaneni KM. Analysing the clinical knowledge of ChatGPT in medical microbiology in the undergraduate medical examination. *Indian J Med Microbi* 2023;45.doi: 10.1016/j.ijmmb.2023.100380
33. Reddy JS, Usha AP, Appavu R, Surapaneni KM. Analyzing the Surgical Knowledge of ChatGPT in Undergraduate Written Medical Examination. *Indian J Surg* 2023.doi: 10.1007/s12262-023-03776-2
34. Sinha RK, Roy AD, Kumar N, Mondal H. Applicability of ChatGPT in Assisting to Solve Higher Order Problems in Pathology. *CUREUS JOURNAL OF MEDICAL SCIENCE* 2023;15(2).doi: 10.7759/cureus.35237
35. Haver HL, Ambinder EB, Bahl M, Oluyemi ET, Jeudy J, Yi PH. Appropriateness of Breast Cancer Prevention and Screening Recommendations Provided by ChatGPT. *Radiology* 2023:230424.doi: 10.1148/radiol.230424

36. Barat M, Soyer P, Dohan A. Appropriateness of Recommendations Provided by ChatGPT to Interventional Radiologists. *Canadian Association of Radiologists Journal* 2023.doi: 10.1177/08465371231170133
37. Potapenko I, Boberg-Ans LC, Hansen MS, Klefter ON, van Dijk E, Subhi Y. Artificial intelligence-based chatbot patient information on common retinal diseases using ChatGPT. *Acta Ophthalmol* 2023.doi: 10.1111/aos.15661
38. Rasmussen M, Larsen AC, Subhi Y, Potapenko I. Artificial intelligence-based ChatGPT chatbot responses for patient and parent questions on vernal keratoconjunctivitis. *Graef Arch Clin Exp* 2023.doi: 10.1007/s00417-023-06078-1
39. Munoz-Zuluaga C, Zhao Z, Wang F, Greenblatt MB, Yang HS. Assessing the Accuracy and Clinical Utility of ChatGPT in Laboratory Medicine. *Clin Chem* 2023.doi: 10.1093/clinchem/hvad058.PubMed: 37231970
40. Samaan JS, Yeo YH, Rajeev N, et al. Assessing the Accuracy of Responses by the Language Model ChatGPT to Questions Regarding Bariatric Surgery. *Obes Surg* 2023.doi: 10.1007/s11695-023-06603-5
41. Yeo YH, Samaan JS, Ng WH, et al. Assessing the performance of ChatGPT in answering questions regarding cirrhosis and hepatocellular carcinoma. *Clinical and molecular hepatology* 2023.doi: 10.3350/cmh.2023.0089
42. Morreel S, Mathysen D, Verhoeven V. Aye, AI! ChatGPT passes multiple-choice family medicine exam. *Med Teach* 2023.doi: 10.1080/0142159X.2023.2187684
43. Balel Y. Can ChatGPT be used in oral and maxillofacial surgery? *Journal of stomatology, oral and maxillofacial surgery* 2023:101471.doi: 10.1016/j.jormas.2023.101471
44. Zhu L, Mou W, Chen R. Can the ChatGPT and other large language models with internet-connected database solve the questions and concerns of patient with prostate cancer and help democratize medical knowledge? *J Transl Med* 2023;21(1).doi: 10.1186/s12967-023-04123-5
45. Schulte B. Capacity of ChatGPT to Identify Guideline-Based Treatments for Advanced Solid Tumors. *Cureus* 2023;15(4):e37938.doi: 10.7759/cureus.37938.PubMed: 37220429
46. Howard A, Hope W, Gerada A. ChatGPT and antimicrobial advice: the end of the consulting infection doctor? *Lancet Infect Dis* 2023;23(4):405-406
47. Ali MJ. ChatGPT and Lacrimal Drainage Disorders: Performance and Scope of Improvement. *Ophthal Plast Recons* 2023;39(3):221-225.doi: 10.1097/IOP.0000000000002418
48. Lee TC, Staller K, Botoman V, Pathipati MP, Varma S, Kuo B. ChatGPT Answers Common Patient Questions About Colonoscopy. *Gastroenterology* 2023.doi: 10.1053/j.gastro.2023.04.033
49. Li SW, Kemp MW, Logan SJS, et al. ChatGPT Outscored Human Candidates in a Virtual Objective Structured Clinical Examination (OSCE) in Obstetrics and Gynecology. *Am J Obstet Gynecol* 2023.doi: 10.1016/j.ajog.2023.04.020
50. Nakhleh A, Spitzer S, Shehadeh N. ChatGPT's Response to the Diabetes Knowledge Questionnaire: Implications for Diabetes Education. *Diabetes Technol the* 2023.doi: 10.1089/dia.2023.0134
51. Ayers JW, Poliak A, Dredze M, et al. Comparing Physician and Artificial Intelligence Chatbot Responses to Patient Questions Posted to a Public Social Media Forum. *Jama Intern Med* 2023.doi: 10.1001/jamainternmed.2023.1838
52. Ayoub NF, Lee YJ, Grimm D, Balakrishnan K. Comparison Between ChatGPT and Google Search as Sources of Postoperative Patient Instructions. *Jama Otolaryngol* 2023.doi: 10.1001/jamaoto.2023.0704
53. Hirosawa T, Harada Y, Yokose M, Sakamoto T, Kawamura R, Shimizu T. Diagnostic Accuracy of Differential-Diagnosis Lists Generated by Generative Pretrained Transformer 3 Chatbot for Clinical Vignettes with Common Chief Complaints: A Pilot Study. *Int J Env Res Pub He* 2023;20(4).doi: 10.3390/ijerph20043378
54. Uz C, Umay E. Dr ChatGPT": Is it a reliable and useful source for common rheumatic diseases? *Int J Rheum*

*Dis* 2023.doi: 10.1111/1756-185X.14749.PubMed: 37218530

55. Seth I, Cox A, Xie Y, et al. Evaluating Chatbot Efficacy for Answering Frequently Asked Questions in Plastic Surgery: A ChatGPT Case Study Focused on Breast Augmentation. *Aesthet Surg J* 2023.doi: 10.1093/asj/sjad140
56. Subramani M, Jaleel I, Krishna Mohan S. Evaluating the performance of ChatGPT in medical physiology university examination of phase I MBBS. *Adv Physiol Educ* 2023;47(2):270-271.doi: 10.1152/advan.00036.2023
57. Alberts IL, Mercolli L, Pyka T, et al. Large language models (LLM) and ChatGPT: what will the impact on nuclear medicine be? *Eur J Nucl Med Mol I* 2023;50(6):1549-1552.doi: 10.1007/s00259-023-06172-w
58. Giannos P, Delardas O. Performance of ChatGPT on UK Standardized Admission Tests: Insights From the BMAT, TMUA, LNAT, and TSA Examinations. *JMIR Med Educ* 2023;9:e47737.doi: 10.2196/47737.PubMed: 37099373
59. Kung TH, Cheatham M, Medenilla A, et al. Performance of ChatGPT on USMLE: Potential for AI-assisted medical education using large language models. *PLOS Digit Health* 2023;2(2):e0000198.doi: 10.1371/journal.pdig.0000198.PubMed: 36812645
60. Cadamuro J, Cabitza F, Debeljak Z, et al. Potentials and pitfalls of ChatGPT and natural-language artificial intelligence models for the understanding of laboratory medicine test results. An assessment by the European Federation of Clinical Chemistry and Laboratory Medicine (EFLM) Working Group on Artificial Intelligence (WG-AI). *Clin Chem Lab Med* 2023.doi: 10.1515/cclm-2023-0355
61. Juhi A, Pipil N, Santra S, Mondal S, Behera JK, Mondal H. The Capability of ChatGPT in Predicting and Explaining Common Drug-Drug Interactions. *CUREUS JOURNAL OF MEDICAL SCIENCE* 2023;15(3).doi: 10.7759/cureus.36272
62. Chervenak J, Lieman H, Blanco-Breindel M, Jindal S. The promise and peril of using a large language model to obtain clinical information: ChatGPT performs strongly as a fertility counseling tool with limitations. *Fertil Steril* 2023.doi: 10.1016/j.fertnstert.2023.05.151.PubMed: 37217092
63. Young JN, O'Hagan R, Poplausky D, et al. The utility of ChatGPT in generating patient-facing and clinical responses for melanoma. *J Am Acad Dermatol* 2023.doi: 10.1016/j.jaad.2023.05.024
64. Lyu Q, Tan J, Zapadka ME, et al. Translating radiology reports into plain language using ChatGPT and GPT-4 with prompt learning: results, limitations, and potential. *VISUAL COMPUTING FOR INDUSTRY BIOMEDICINE AND ART* 2023;6(1).doi: 10.1186/s42492-023-00136-5
65. Johnson SB, King AJ, Warner EL, Aneja S, Kann BH, Bylund CL. Using ChatGPT to evaluate cancer myths and misconceptions: artificial intelligence and cancer information. *JNCI CANCER SPECTRUM* 2023;7(2).doi: 10.1093/jncics/pkad015
66. Van Bulck L, Moons P. What if your patient switches from Dr. Google to Dr. ChatGPT? A vignette-based survey of the trustworthiness, value, and danger of ChatGPT-generated responses to health questions. *Eur J Cardiovasc Nur* 2023.doi: 10.1093/eurjcn/zvad038
67. Duong D, Solomon BD. Analysis of large-language model versus human performance for genetics questions. *Eur J Hum Genet* 2023.doi: 10.1038/s41431-023-01396-8.PubMed: 37246194
68. Huh S. Are ChatGPT's knowledge and interpretation ability comparable to those of medical students in Korea for taking a parasitology examination?: a descriptive study. *JOURNAL OF EDUCATIONAL EVALUATION FOR HEALTH PROFESSIONS* 2023;20.doi: 10.3352/jeehp.2023.20.1
69. Aldridge MJ, Penders R. Artificial intelligence and anaesthesia examinations: exploring ChatGPT as a prelude to the future. *Br J Anaesth* 2023.doi: 10.1016/j.bja.2023.04.033.PubMed: 37244834
70. Shay D, Kumar B, Bellamy D, et al. Assessment of ChatGPT success with specialty medical knowledge using

anaesthesiology board examination practice questions. *Brit J Anaesth* 2023.doi: 10.1016/j.bja.2023.04.017

71. Fijaoko N, Gosak L, Stiglic G, Picard CT, Douma MJ. Can ChatGPT pass the life support exams without entering the American heart association course? *Resuscitation* 2023;185.doi: 10.1016/j.resuscitation.2023.109732
72. Suchman K, Garg S, Trindade AJ. ChatGPT Fails the Multiple-Choice American College of Gastroenterology Self-Assessment Test. *Am J Gastroenterol* 2023.doi: 10.14309/ajg.0000000000002320.PubMed: 37212584
73. Oh N, Choi GS, Lee WY. ChatGPT goes to the operating room: evaluating GPT-4 performance and its potential in surgical education and training in the era of large language models. *Ann Surg Treat Res* 2023;104(5):269-273.doi: 10.4174/astr.2023.104.5.269.PubMed: 37179699
74. Humar P, Asaad M, Bengur FB, Nguyen V. ChatGPT is Equivalent to First Year Plastic Surgery Residents: Evaluation of ChatGPT on the Plastic Surgery In-Service Exam. *Aesthet Surg J* 2023.doi: 10.1093/asj/sjad130
75. Deebel NA, Terlecki R. ChatGPT performance on the American Urological Association (AUA) Self-Assessment Study Program and the potential influence of artificial intelligence (AI) in urologic training. *Urology* 2023.doi: 10.1016/j.urology.2023.05.010
76. Hopkins BS, Nguyen VN, Dallas J, et al. ChatGPT versus the neurosurgical written boards: a comparative analysis of artificial intelligence/machine learning performance on neurosurgical board-style questions. *J Neurosurg* 2023:1-8.doi: 10.3171/2023.2.JNS23419
77. Gilson A, Safranek CW, Huang T, et al. How Does ChatGPT Perform on the United States Medical Licensing Examination? The Implications of Large Language Models for Medical Education and Knowledge Assessment. *JMIR Med Educ* 2023;9:e45312.doi: 10.2196/45312.PubMed: 36753318
78. Mihalache A, Popovic MM, Muni RH. Performance of an Artificial Intelligence Chatbot in Ophthalmic Knowledge Assessment. *Jama Ophthalmol* 2023.doi: 10.1001/jamaophthalmol.2023.1144
79. Bhayana R, Krishna S, Bleakney RR. Performance of ChatGPT on a Radiology Board-style Examination: Insights into Current Strengths and Limitations. *Radiology* 2023:230582.doi: 10.1148/radiol.230582
80. Wang YM, Shen HW, Chen TJ. Performance of ChatGPT on the Pharmacist Licensing Examination in Taiwan. *J Chin Med Assoc* 2023.doi: 10.1097/JCMA.0000000000000942.PubMed: 37227901
81. Gupta R, Herzog I, Park JB, et al. Performance of ChatGPT on the Plastic Surgery Inservice Training Examination. *Aesthet Surg J* 2023.doi: 10.1093/asj/sjad128
82. Thirunavukarasu AJ, Hassan R, Mahmood S, et al. Trialling a Large Language Model (ChatGPT) in General Practice With the Applied Knowledge Test: Observational Study Demonstrating Opportunities and Limitations in Primary Care. *JMIR Med Educ* 2023;9:e46599.doi: 10.2196/46599.PubMed: 37083633
83. Spellberg B, Harrington D, Black S, Sue D, Stringer W, Witt M. Capturing the diagnosis: an internal medicine education program to improve documentation. *The American Journal of Medicine* 2013;126(8):739-743
84. Savoia P. Skills, Knowledge, and Status:: The Career of an Early Modern Italian Surgeon. *B Hist Med* 2019;93(1):27
85. Wolff RF, Moons K, Riley RD, et al. PROBAST: A Tool to Assess the Risk of Bias and Applicability of Prediction Model Studies. *Ann Intern Med* 2019;170(1):51-58.doi: 10.7326/M18-1376.PubMed: 30596875
86. Higgins JP, Altman DG, Gøtzsche PC, et al. The Cochrane Collaboration's tool for assessing risk of bias in randomised trials. *BMJ* 2011;343:d5928.doi: 10.1136/bmj.d5928.PubMed: 22008217
87. Slim K, Nini E, Forestier D, Kwiatkowski F, Panis Y, Chipponi J. Methodological index for non-randomized studies (minors): development and validation of a new instrument. *Anz J Surg* 2003;73(9):712-6.doi: 10.1046/j.1445-2197.2003.02748.x.PubMed: 12956787
88. Stang A. Critical evaluation of the Newcastle-Ottawa scale for the assessment of the quality of nonrandomized studies in meta-analyses. *Eur J Epidemiol* 2010;25(9):603-5.doi: 10.1007/s10654-010-9491-

z.PubMed: 20652370

## Table 1. Evaluation Framwork for Large Language Models in Medicine

| Section | Item | Checklist Item | Report |
|---|---|---|---|
| Task Generation | 1 | Source of the task: e.g., website, clinical vignette, exam question bank, book | |
| | 2 | Source date | |
| | 3 | Number (of the question/task) | |
| LLM Version | 4 | LLM version (e.g., GPT-4) | |
| | 5 | Date of inquiry | |
| Conversation Structure | 6 | Prompt of the inquiry (e.g., "Please answer..." or "Please role-play as a doctor and respond...") | |
| | 7 | Mode of inquiry: on-page or API call | |
| | 8 | Are the questions individual standalone queries or a continuous conversation requiring multiple consecutive inquiries? | |
| | 9 | Is the inquiry independent? (e.g., new chat) | |
| | 10 | Language (e.g., English, Chinese, Dutch) | |
| | 11 | Repetition of the inquiry: Is the inquiry repeated, and if so, how many repetitions? | |
| Evaluation | 12 | Who is the evaluator? (e.g., Internal medicine doctor, Pediatrician, Radiologist) | |
| | 13 | Is the expert assessor blinded? Is the expert assessor unaware of whether they are conversing with a human or an AI? | |
| | 14 | Number of evaluators | |
| | 15 | Evaluation metrics (e.g., Accuracy, Completeness, Safety, Empathy, Text length) | |
| | 16 | Does the response address the core question of the conversation | |
| | 17 | Overall response evaluation: the agreed evalutaion metrics should be used to give an overall evaluation for the response | |
| | 18 | Is quantitative evaluation employed? If yes, what are the quantitative metrics? (e.g., Likert scale, Accuracy rate) | |
| | 19 | If repeated inquiries, do repeated questions of the same query to the algorithm (or to different human experts) produce substantially different replies? How consistent are the responses? | |
| | 20 | If multiple evaluators, is consistency evaluated? | |

LLM: large language model; API: Application Programming Interface; AI: artificial intelligence

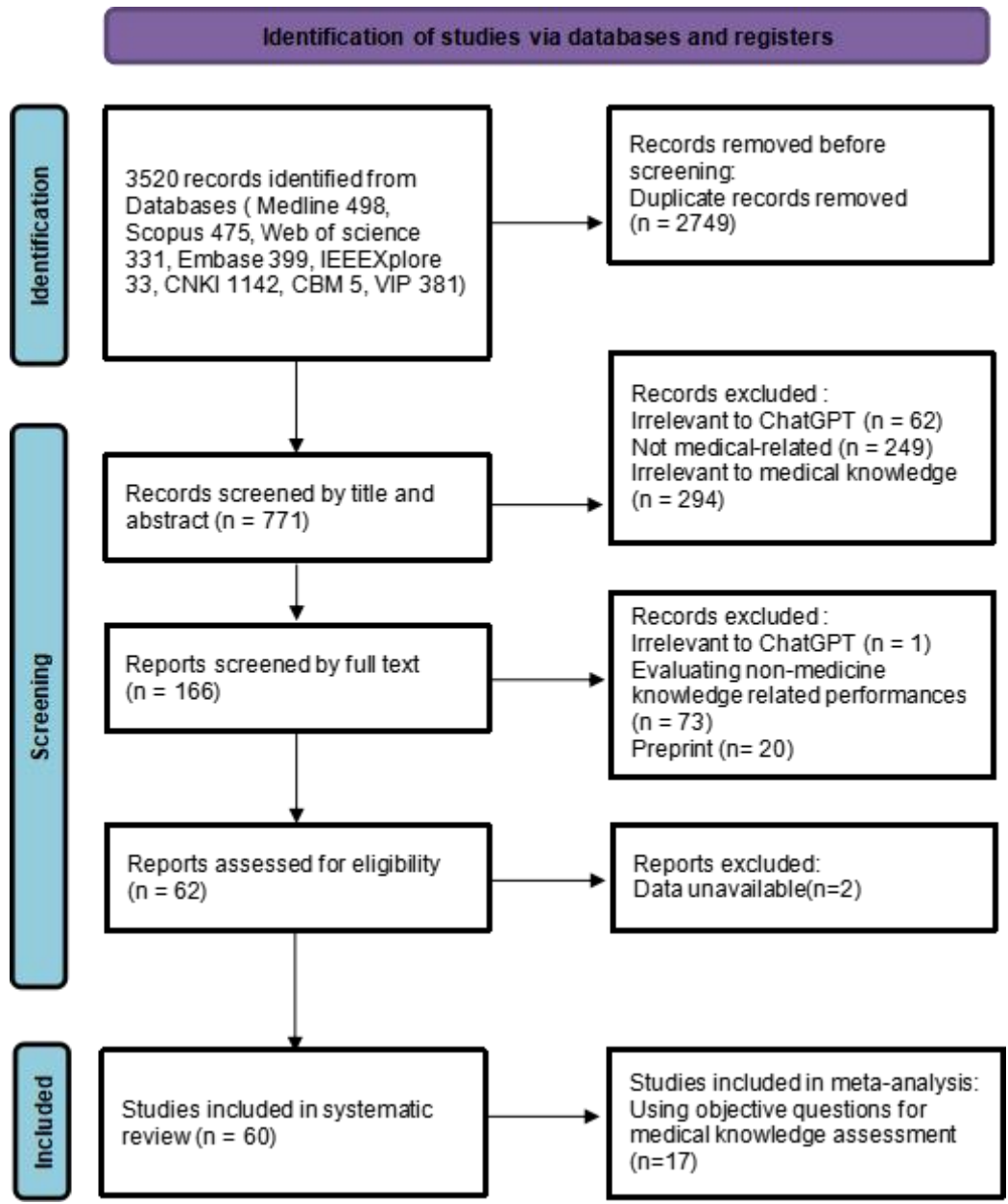

**Figure 1.** Flowchart for the literature screening and selection process. CNKI: China National Knowledge Infrastructure; CBM: Chinese BioMedical Literature Database; VIP: VIP Database for Chinese Technical Periodicals

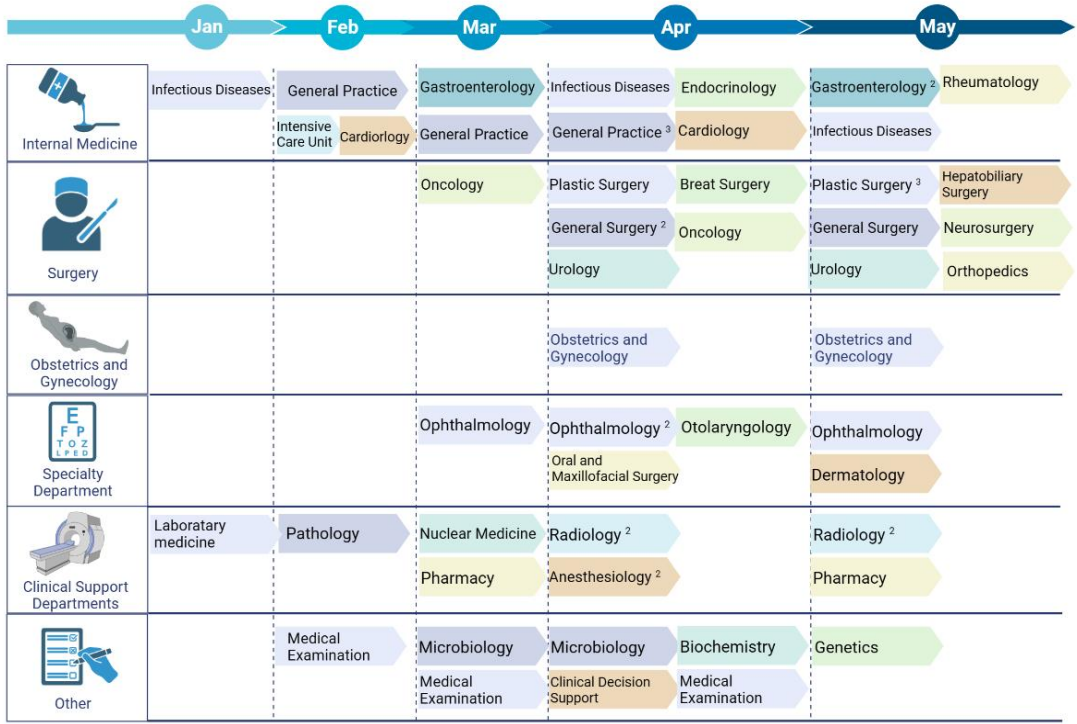

**Figure 2.** Medical Examinations and Assessments Undertaken by ChatGPT

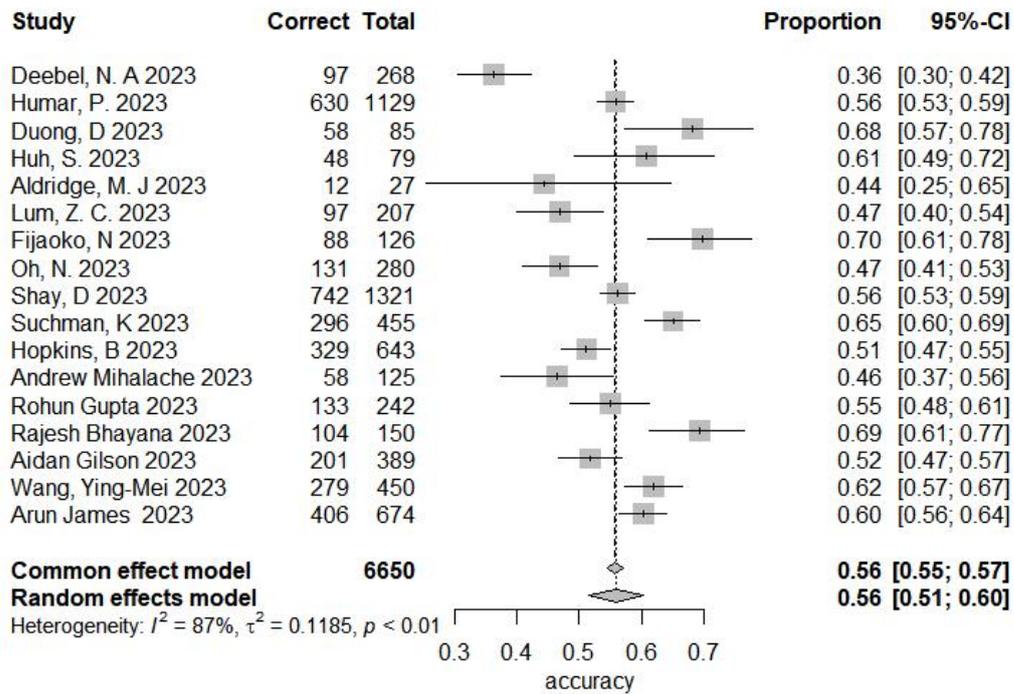

**Figure 3.** Performance of ChatGPT. Correct: the number of correctly answered questions; Total: the total number of questions asked; 95%CI: 95% Confidence Interval; Common effect: Fixed-effect